# A model to explain the Q-increase by moderate-temperature treatment in Nb SRF cavities

Yegor Tamashevich[1], Alena Prudnikava[1] and Jens Knobloch[1,2]

[1] Helmholtz Centre for Materials and Energy, D-12489 Berlin, Germany
[2] Department of Physics, Universität Siegen, Walter-Flex-Str. 3, 57068 Siegen, Germany

E-mail: yegor.tamashevich@helmholtz-berlin.de

**Abstract**

It is well known that moderate temperature baking of niobium cavities can improve the surface resistance. Presently, it is believed that the diffusion of oxygen into the bulk, resulting in interstitial defects, is responsible for the change. In this note we propose that the damaged surface layer remaining after dissolution of the thin niobium pentoxide may in fact be the dominant contributor to the improved the cavity quality factor by strongly pinning trapped flux lines. We propose some of experiments to test this theory.

## 1. Impact of moderate-temperature heat treatment on the RF surface

Currently, the increase of the intrinsic quality $Q_0$ factor of Nb SRF cavities after heat treatment at moderate temperatures (120-400 °C) is attributed to a specific depth profile of oxygen interstitials in niobium lattice. During the heat treatment, the niobium pentoxide layer "dissolves", and the released oxygen atoms diffuse into the bulk, forming an "oxygen-rich" layer. The oxygen interstitials are basically lattice defects and believed to increase the $Q_0$ by a number of potential mechanisms: decrease of the mean free path of the normal-conducting electrons, thus decreasing BCS resistance $R_{BSC}$, enhanced flux pinning to reduce the residual resistance $R_{res}$, or by acting as hydrogen trapping centers to avoid hydride precipitation. Most of the recent studies measure and describe this profile of oxygen concentration and look for an optimal treatment procedure (temperature, time etc.) to tailor this profile.

However, we believe an additional mechanism, plays a role. The distorted crystal lattice at the surface of niobium left after the dissolution of the niobium pentoxide during heat treatment may in fact dominate the Q-increase. A qualitative explanation is shown in Figure 1.

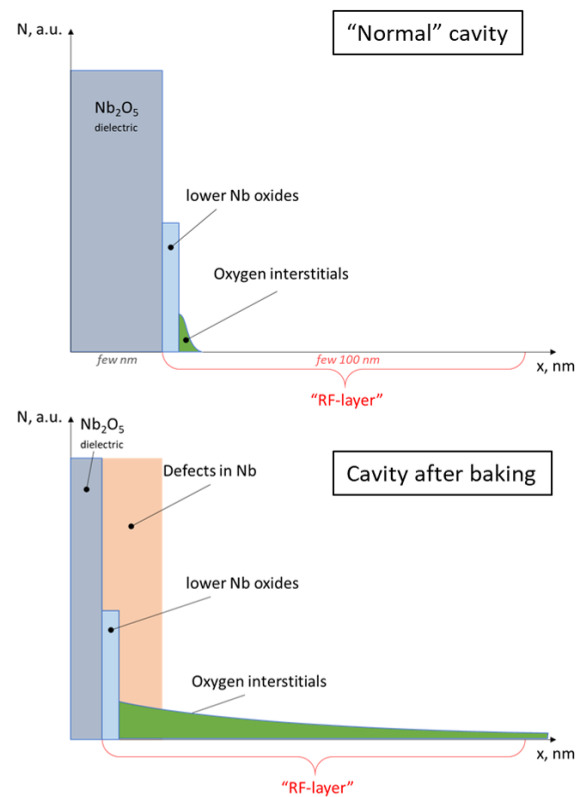

*Figure 1. Schematic of the RF surface in superconducting Nb cavities before and after mid-T heat treatment. (Note: lengths are not to scale; the RF layer is much deeper than the pentoxide thickness).*

In the top picture, the concentration of oxygen in the near-surface layer of an unbaked cavity is shown. It consists of a natural niobium pentoxide ($Nb_2O_5$), some Nb oxides of lower oxidation states (shown simplified, see [1] for details), oxygen interstitials and bulk niobium underneath. The pentoxide layer, being dielectric, does not contribute to rf losses. Thus, most of the superconducting current flows below the pentoxide in the layer limited to ca. three times the London penetration depth. Assuming a standard 800 °C annealed cavity, this layer consists of almost perfect niobium with some amount of oxygen atoms in a thin layer of lower niobium oxides right below the pentoxide layer. The pentoxide layer has thickness of ca. 3.5-5 nm, the "lower Nb oxides" layer has the thickness of ca. 1 nm, and the "RF-layer" has the thickness of ca. 150 nm. The oxygen concentration in the pentoxide layer is ca. 70 at. %, in the simplified "lower Nb oxides" layer – ca. 40-50 at. %. For the simplicity, we can assume that initially are no other interstitials or defects in the RF layer.

The bottom picture represents the concentration of oxygen-interstices and distorted Nb-lattice in the near-surface layer of a cavity, which was heat-treated at moderate temperature. The pentoxide layer is partially or completely dissociated, and the released oxygen diffuses into bulk niobium forming a profile of a decaying concentration. During the baking process the inner border of the pentoxide phase effectively moves towards the surface but, importantly, the lattice distortion remains even in the dissociated region because the temperature is too low to recrystallize the niobium. What remains is a metallic region with a very high defect concentration. The rate of dissociation and diffusion is strongly temperature-dependent. Thus, the pentoxide thickness and the concentration profile of the oxygen interstitials can be tailored by adjusting the baking time and temperature. In general, if the treatment is performed for several hours at moderate temperatures (300-400 °C), oxygen-interstitials profile extends beyond the "RF-layer" [2, 3, 4]. Oxide dissociation and diffusion is negligible at temperatures below ca. 120 °C. Thus, we assume that the oxygen profile remains constant after the heat treatment and the cavity is cooled down to room temperature or below.

## 2. Model for reduced surface resistance

The key point of the present note is that a distorted niobium lattice remains under the pentoxide in the region where the original pentoxide dissolved. All the vacancy-defects left after the oxygen has diffused out of the pentoxide layer stay in-place and are not "cured". The concentration of these defects is very high, reaching almost 70 at %, i.e., the Nb lattice in this layer is severely damaged. The total number of these defects equals the total number of oxygen interstitials remaining in the bulk. However, since the profile may extend beyond the rf layer, the defects from the pentoxide near the surface may in fact well outnumber the oxygen interstitials *within the RF layer*. Furthermore, the concentration of these near-surface defects will be much higher than the concentration of interstitials deeper in the RF layer. As a result, we believe enhanced flux pinning at the surface will inhibit flux oscillation in the RF field and hence reduce the residual resistance. In light of the above, the layer of defects right below the pentoxide layer should have a significant (or even a dominant) contribution to the Q-increase phenomena.

## 3. Some consequences and tests of the model

Several consequences of the described hypothesis can be deduced. Examples include:

1) The originally increased quality factor upon baking should continuously decrease if a cavity is subsequently subjected to air, thereby regrowing the original pentoxide layer. The newly formed pentoxide is dielectric and is not a part of the RF-layer. The number of defects in RF layer decreases and hence the surface resistance increases again. This process may require several days or even weeks before the original pentoxide thickness is re-established. A series of sequential tests of a heat-treated cavity subjected to air should demonstrate a gradual degradation of the quality factor. After the pentoxide layer is fully formed, any remaining improvement of the surface resistance (compared to untreated cavity) can be then attributed to the oxygen interstitials in the RF layer.

2) In the absence of any trapped flux, the surface resistance of the distorted lattice will depend on its thickness and defect concentration. Hence, including trapped flux there should be an optimal concentration of defects to produce a minimum of residual resistance (see Figure 2). An untreated cavity with few defects will likely have a non-minimum residual resistance (as well as BCS resistance). As the defect concentration increases due to heat treatment, flux motion is reduced by trapping, and the residual surface resistance drops. However, the residual resistance due to other effects increases and for prolonged heat treatment may eventually (at very high defect concentration) dominate. Hence a minimum surface resistance (optimal heat treatment time/temperature) is likely. Successively longer heat treatments should therefore result in the residual resistance evolving similar to Figure 2. The optimum must be found experimentally.

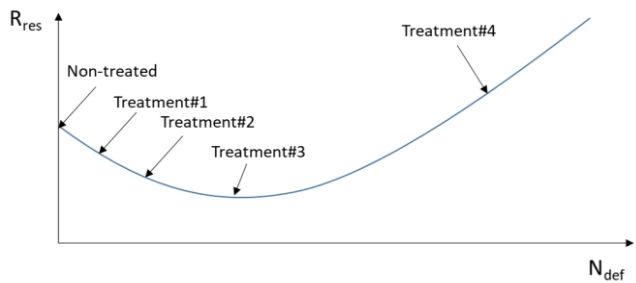

*Figure 2: Expected evolution of the residual resistance as a function of near surface defects for increasingly long heat treatments (or at higher temperatures).*

3) If strong flux pinning is responsible for the reduced residual resistance, then a reduction in sensitivity ($dR_{res}/dB$) should be observed following mid-T baking. This can be measured by applying varying external magnetic fields during prior to cavity testing.

4) Niobium with good flux expulsion characteristics (e.g., single-crystal Nb) should decrease its ability to expel flux when the pentoxide layer is reduced by moderate-temperature baking due to the increased trapping centers. Exposure to air should then re-establish the original flux expulsion performance. Such tests do not require cavities, but can be performed on samples (see e.g. [5, 6]).

5) The maximum number of defects is limited by the original thickness of the pentoxide layer. All the defects can be utilised only if the full pentoxide layer is dissociated. However, in that case the surface has no protective layer and a significant contamination can occur (especially with carbon) (see [1]). If the initial pentoxide layer can be made thicker (for example, by anodic oxidation), more defects can be introduced in to the rf layer. Currently we do not know if the total amount of defects in a natural pentoxide layer is enough to reach the minimum of $R_{res}$. It is possible, that with the natural pentoxide we can reach, for example, Treatment#2 values in Figure 2.

## References

[1] A. Prudnikava 2023. HZB experimental report. To be published in July 2023.
[2] C. Y. Ang, "Activation energies and diffusion coefficients of oxygen and nitrogen in niobium and tantalum." Acta Metallurgica 1.2 (1953): 123-125.
[3] F. J. M. Boratto and R. E. Reed-Hill. "On the calculations of the diffusion coefficients of oxygen and nitrogen in niobium." Metallurgical Transactions A 8 (1977): 1233-1238.
[4] R. W. Powers and Margaret V. Doyle. "Diffusion of interstitial solutes in the group V transition metals." Journal of Applied Physics 30.4 (1959): 514-524.
[5] F. Kramer, "Impact of Cooldown Conditions on Trapped Flux in Superconducting Niobium", PhD Thesis, Universität Siegen, 2023.
[6] F. Kramer, invited talk, SRF2023 (to be published).